\documentclass[12pt,a4paper, amssymb, amsmath]{article}

\usepackage{achemso}
\usepackage{epsfig}
\usepackage{graphicx}
\usepackage{indentfirst}
\usepackage{multirow}
\usepackage{latexsym}
\title{{How does quantum confinement influence the electronic structure of transition metal sulfides TmS$_2$}}

\author{A.~Kuc, N.~Zibouche and T.~Heine \\
School of Engineering and Science, Jacobs University Bremen, \\
Campus Ring 1, 28759 Bremen, Germany}

\date{\today}
\begin{document}
\maketitle

\begin{abstract}

Bulk MoS$_2$, a prototypical layered transition-metal dichalcogenide, is an indirect band gap semiconductor.
Reducing its size to a monolayer, MoS$_2$ undergoes a transition to the direct band semiconductor.
We support this experimental observation by first principles calculations and show that quantum confinement in layered $d$-electron dichalcogenides results in tuning the electronic structure at the nanoscale.
We further studied the properties of related TmS$_2$ nanolayers (Tm = W, Nb, Re) and show that the isotopological WS$_2$ exhibits similar electronic properties, while NbS$_2$ and ReS$_2$ remain metallic independent on size.

\end{abstract}

\section{Introduction}

\begin{figure}[ht!]
\begin{center}
\includegraphics[scale=0.20,clip]{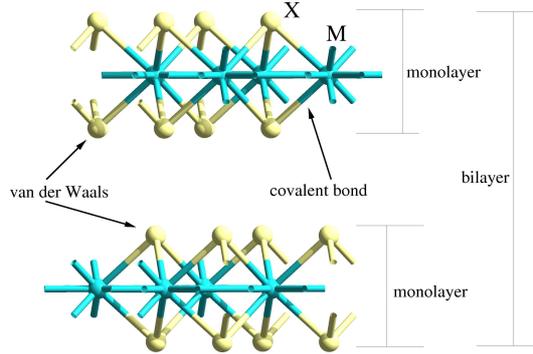}
\caption{\label{fig:1}The atomic structure of layered transition-metal dichalcogenides of TmX$_2$ type (Tm -- transition metal, X -- chalcogenide). Different sheets of TmX$_2$ are composed of three atomic layers X-Tm-X, where Tm and X are covalently bonded. Sheets are held together by weak van der Waals forces.}
\end{center}
\end{figure}

Layered transition-metal dichalcogenides (LTMDCs) of TmX$_2$ type (Tm = Mo, W, Nb, Re, Ti, Ta, etc., X = S, Se, Te, etc.) have been studied extensively on the experimental and theoretical basis for the last 40 years.\cite{Wilson1969, Mattheis1973, Mattheis1973a, Kam1982, Coehoorn1987, Coehoorn1987a, Kobayashi1995, Wilcoxon1997, Reshak2005, Lebegue2009, Arora2009, Splendiani2010, Mak2010, Li2010, Heda2010, Matte2010, Liu2010}
Molybdenum disulfide (MoS$_2$) is a prototypical LTMDC, which is composed of two-dimensional S-Mo-S sheets stacked on top of one another, as shown in Fig.~\ref{fig:1}.
Each sheet is trilayered with a Mo atom in the middle that is covalently bonded to six S atoms located in the top and bottom layers.
The bonding between the adjacent S-Mo-S sheets is much weaker and they are held together by van der Waals forces leading to the two-dimensional character of MoS$_2$.

Due to the weak forces between the sheets and the anisotropic character of LTMDCs, shearing takes place more easily even under high pressure, leading to lubricant applications.\cite{Mattheis1973a, Drummond2001}
Other applications, such as catalysis, optoelectronics and photovoltaics, have been proposed and investigated as well.\cite{Wilson1969, Kam1982, Tenne1985, Coehoorn1987, Sienicki1996, Gourmelon1997}
A single layer of MoS$_2$ was recently used to realize a field-effect transistor with HfO$_2$ as a gate insulator.\cite{Radisavljevic2011}
The transistor exhibits a room-temperature current on/off ratio exceeding 1x10$^8$ and mobility comparable to the mobility of thin silicon films or graphene nanoribbons.
Two-dimensional sheets of layered materials can be produced nowadays by e.g.\ liquid exfoliation, what was successfully performed in case of transition-metal dichalcogenides by Coleman et al.\cite{Coleman2011} 

Electronic structure of various bulk TMDCs and a monolayer of MoS$_2$ has been previously studied throughout $ab~initio$ calculations using plane waves approach or local basis functions.\cite{Kobayashi1995, Reshak2005, Lebegue2009, Splendiani2010, Li2010}
MoS$_2$ is an indirect band gap material in its bulk form, which recently was shown to become a direct band gap semiconductor when thinned to a monolayer.\cite{Li2007, Splendiani2010, Matte2010}
Therefore, few-layer-MoS$_2$ materials can have their band gaps tuned and their values shift from the bulk value by various amounts due to quantum size effects.

In this paper, we have studied size-dependent electronic properties of TmS$_2$ type, where Tm = Mo, W, Re and Nb.
The first principle calculations were performed using localised Gaussian basis functions and compared to the available experimental data and plane wave calculations.
The results show that the indirect-direct band gap transitions also hold for WS$_2$.
The band structures of ReS$_2$ and NbS$_2$ reveal metallic character and therefore, no transition appears.
It can be expected that the size-dependent phenomenon of band gap transition will occur also for other dichalcogenides, like MoSe$_2$, WSe$_2$, MoTe$_2$ or WTe$_2$.
This study is in progress and will be the subject of next communication.

\section{Methods}
\label{Sec:Methods}

In this work, we have studied LTMDCs of TmS$_2$ type, where Tm = Mo, W, Re and Nb.
All structures have hexagonal symmetry and belong to the $P6_3/mmc$ space group.
ReS$_2$ is a triclinic system but for the purposes of a direct comparison we have used hexagonal symmetry.
The monolayers and polylayers were cut out from the fully optimised bulk structures as (0 0 1) surfaces.
Different numbers of layers were considered in this studies: mono-, bi-, and quadrilayers, as well as, 6- and 8-layered structures.

First-principle calculations were performed on the basis of density functional theory (DFT) as implemented in the CRYSTAL09 code.\cite{CRYSTAL09}
The exchange and correlation terms were described using general gradient approximation (GGA) in the scheme of PBE (Perdew-Burke-Ernzerhof)\cite{PBE} and PBE hybrid (PBE0) functionals.\cite{PBE0}
The following basis sets were used: Mo$\_$SC$\_$HAYWSC-311(d31)G$\_$cora$\_$1997 (for Mo atoms), W$\_$cora$\_$1996 (for W atoms), Nb$\_$SC$\_$HAYWSC-31(d31)G$\_$dallolio$\_$1996 (for Nb atoms), Re$\_$cora$\_$1991 (for Re atoms), and S$\_$86-311G*$\_$lichanot$\_$1993 (for S atoms).

Optimisation of initial experimental structures was performed using analytical energy gradients with respect to atomic coordinates and unit cell parameters within a quasi-Newton scheme combined with the BFGS (Broyden-Fletcher-Goldfarb-Shanno) scheme for Hessian updating.
The optimised lattice parameters for all the studied materials are given in Table~\ref{tab:1}.
\begin{table}
\caption{\label{tab:1}Calculated and experimental lattice parameters of hexagonal transition-metal dichalcogenides in the form of TmX$_2$ (Tm = Mo, W, Nb, Re; X = S, Se). Results obtained at the DFT/PBE level. In parenthesis, data obtained at the DFT/PBE0 level are given. }
\begin{tabular}{c|c|c|c|c}
\multirow{2}{*}{\textbf{~Structure~}} & \multicolumn{2}{c|}{\textbf{Theory}} & \multicolumn{2}{c}{\textbf{Exp.\cite{Wilson1969, Mattheis1973, Coehoorn1987a}}} \\
\cline{2-5}
                   & \textbf{$a$} & \textbf{$c$} & \textbf{$a$} & \textbf{$c$} \\
\hline
\textbf{MoS$_2$} &~~3.173 (3.143)~~&~~12.696 (12.583)~~&~~3.160~~&~~12.295~~\\
\textbf{WS$_2$}  &~~3.164 (3.139)~~&~~12.473 (12.380)~~&~~3.154~~&~~12.362~~\\
\textbf{NbS$_2$} &~~3.332 (3.313)~~&~~12.106 (12.074)~~&~~3.310~~&~~11.890~~\\
\textbf{ReS$_2$} &~~3.300 (3.275)~~&~~12.724 (12.148)~~&~~--   ~~&~~--    ~~\\
\end{tabular}
\end{table}

The shrinking factor for bulk and layered structures was set to 8, what results in the corresponding number of 50 and 30 k-points in the irreducible Brillouin zone, respectively.
The mesh of k-points was obtained according to the scheme proposed by Monkhorst and Pack.\cite{Monkhorst1976}
Band structures were calculated along the high symmetry points using the following path $\it{\Gamma-M-K-\Gamma}$.

\section{Results and Discussion}

\begin{figure}[ht!]
\begin{center}
\includegraphics[scale=0.65,clip]{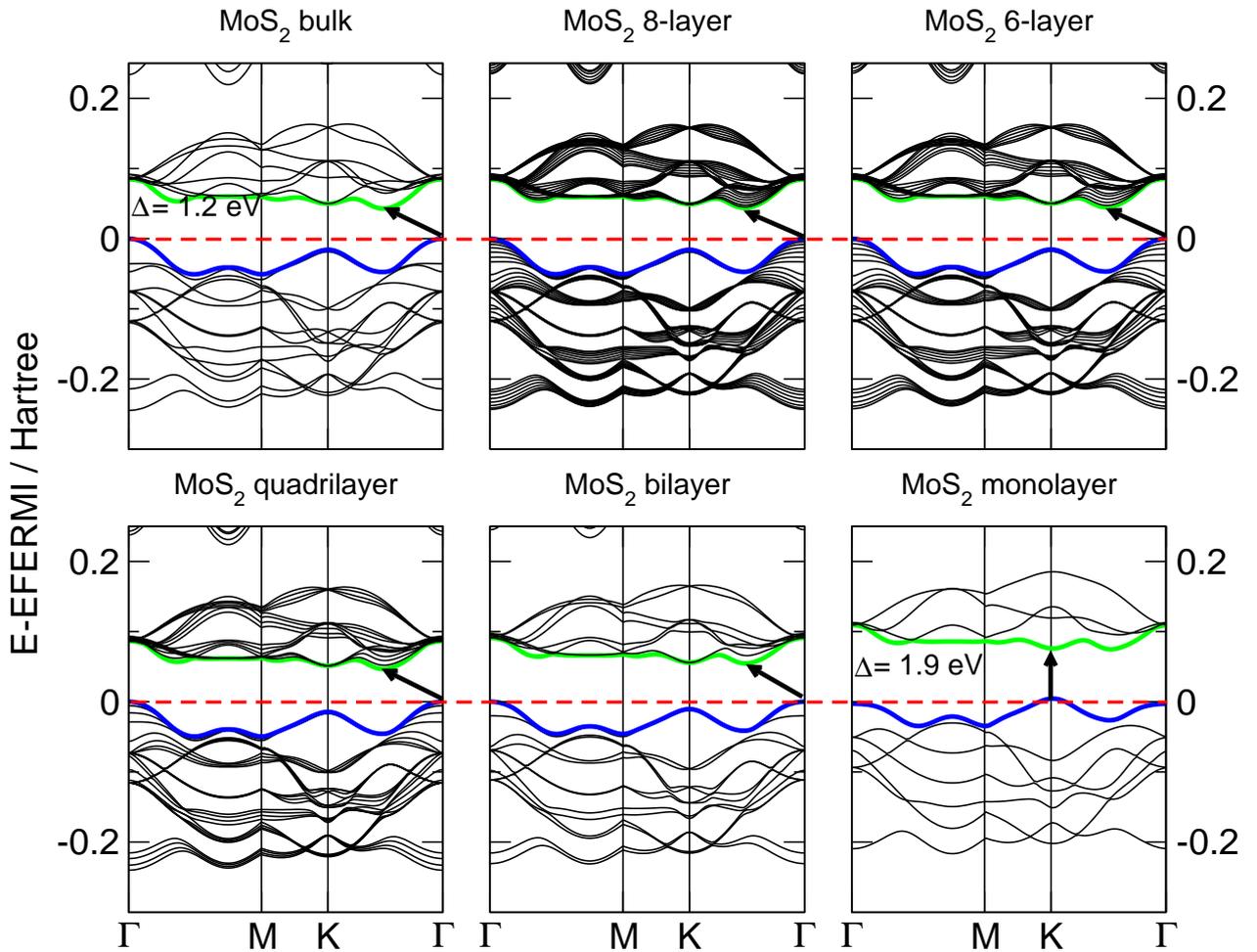}
\caption{\label{fig:2}Band structures of bulk MoS$_2$, its monolayer, as well as, polylayers calculated at the DFT/PBE level. The horizontal dashed lines indicate the Fermi level. The arrows indicate the smallest value of the band gap (direct or indirect) for a given system. The top of valence band (blue) and bottom of conduction band (green) are highlighted (Online color).}
\end{center}
\end{figure}

\begin{figure}[ht!]
\begin{center}
\includegraphics[scale=0.65,clip]{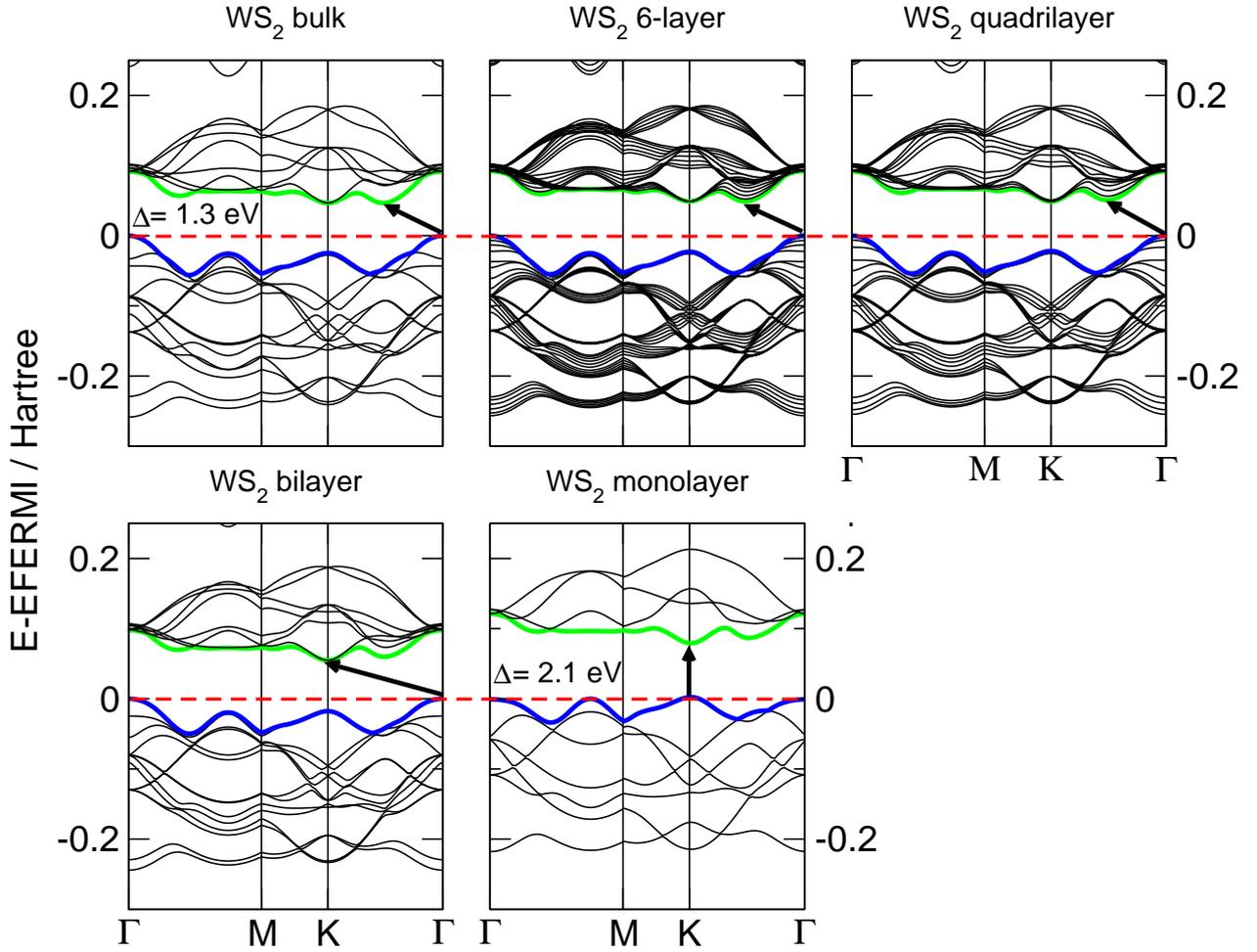}
\caption{\label{fig:3}Band structures of bulk WS$_2$, its monolayer, as well as, polylayers calculated at the DFT/PBE level. The horizontal dashed lines indicate the Fermi level. The arrows indicate the smallest value of the band gap (direct or indirect) for a given system. The top of valence band (blue) and bottom of conduction band (green) are highlighted (Online color).}
\end{center}
\end{figure}

We have studied electronic properties of TmS$_2$ layered structures with respect to the size of the systems.
Bulk structures as well as mono- and polylayers were considered.
Figs.~\ref{fig:2} and \ref{fig:3} show band structures of MoS$_2$ and WS$_2$, respectively, calculated using PBE functional and going from bulk to a monolayer.
[For band structures results using PBE0 functional see Supporting Information].

\begin{figure}[ht!]
\begin{center}
\includegraphics[scale=0.65,clip]{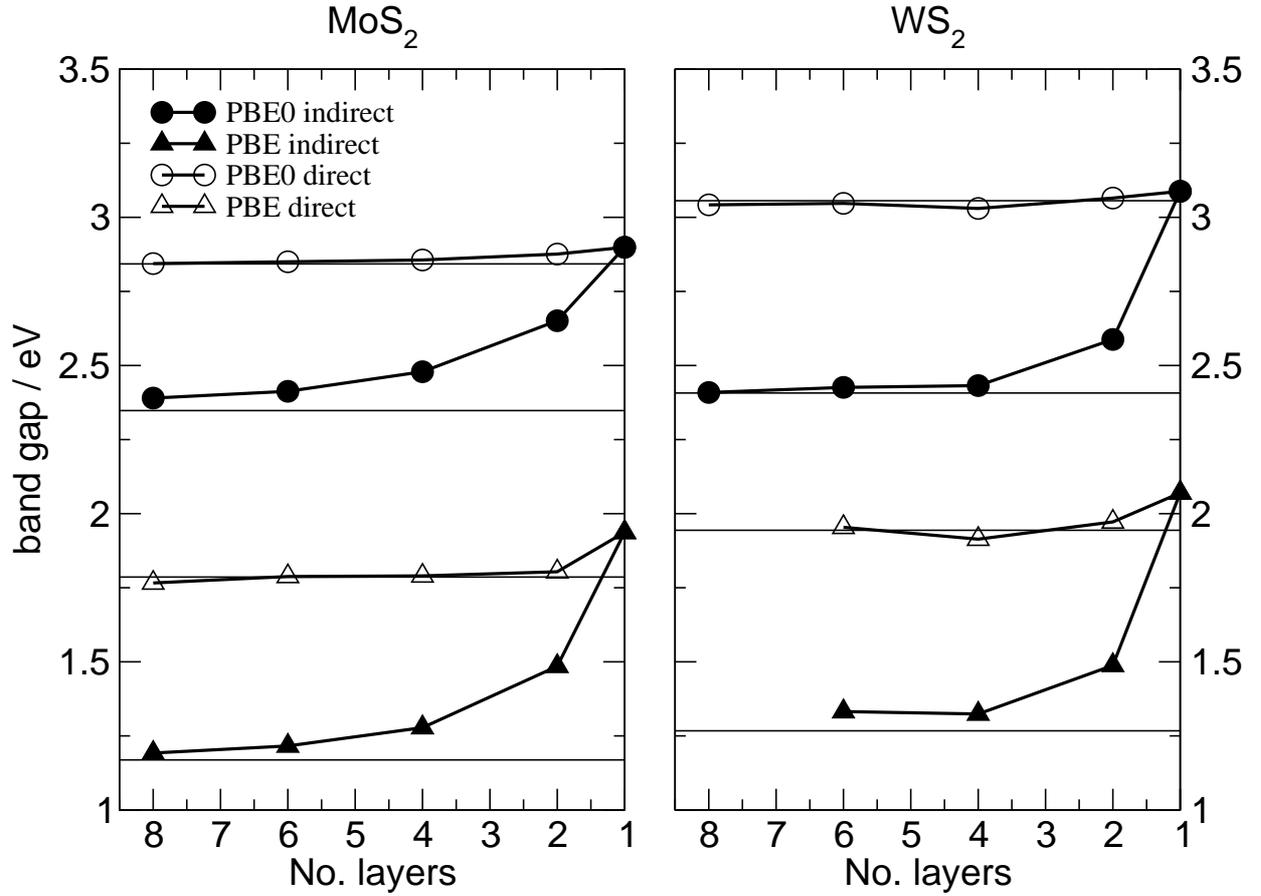}
\caption{\label{fig:4}Calculated direct and indirect band gap values of MoS$_2$ and WS$_2$. The horizontal solid lines indicate the band gaps of bulk structures.}
\end{center}
\end{figure}

The results show that the bulk MoS$_2$ and WS$_2$ are indirect-gap semiconductors.
The band gap originates from transition from the top of valence band situated at $\it{\Gamma}$ to the bottom of conduction band halfway between $\it{\Gamma}$ and $K$ high symmetry points.
The optical direct band gap is situated at $K$ point.
As the number of layers decreases the indirect band gap increases and becomes so high in the monolayer that the material changes into a 2D direct band gap semiconductor.
At the same time, the direct gap stays almost unchanged (independent of size) and close to the value of the direct band gap of a bulk system.
These results are visualised better in Fig.\ref{fig:4}, where the band gap of MoS$_2$ and WS$_2$ are plotted against the number of layers.

There is a significant difference in the band gap values calculated using PBE and PBE0 functionals.
Comparing the results with the available experimental data, one can see that the PBE functional performs  better.
The indirect band gap of bulk MoS$_2$ is 1.23~eV\cite{Kam1982} and our value of 1.2~eV agrees perfectly at the PBE level, while it is overestimated by around 1~eV using PBE0 hybrid functional.
Similarly, the band gap of WS$_2$, with experimental value of 1.35~eV\cite{Kam1982}, is 1.3~eV at the PBE level and 2.4~eV using PBE0.
The experimental direct band gaps of MoS$_2$ and WS$_2$ are 1.74~eV and 1.79~eV, respectively.\cite{Kam1982}
Our DFT/PBE calculations give values of 1.8~eV and 1.9~eV for MoS$_2$ and WS$_2$, respectively.
Again, the PBE0 results are by around 1~eV overestimated.

Li et al.,\cite{Li2010} using GGA/PBE functional and the plane wave approach, obtained band gap of 0.79 eV for the bulk MoS$_2$.
For the same system, Matte et al.\cite{Matte2010} have obtained 1.1~eV and 1.70~eV for the indirect and direct band gap, respectively, using GGA/PBE and DZP basis set.
Using Perdew-Wang exchange-correlation functional and self-consistent pseudopotential, Arora et al.\cite{Arora2009} have obtained the indirect band gap of 1.32~eV for the WS$_2$ bulk.

Decreasing the number of layers causes progressive shift in the indirect gap up to 1.9~eV and 2.1~eV for MoS$_2$ and WS$_2$, respectively.
The change in the indirect gap energy is significantly larger than that of the direct gap, which increased by around 0.1~eV.
Similar results were found by Mak at al.\cite{Mak2010} for MoS$_2$.

The band gaps of bulk structures are in the range of near infrared.
Reducing the size of the material to just a few layers causes the blue shift in absorption features and shifts the band gap to the range of the visible light.
These materials can be, therefore, interesting in optoelectronics.

\begin{figure}[ht!]
\begin{center}
\includegraphics[scale=0.65,clip]{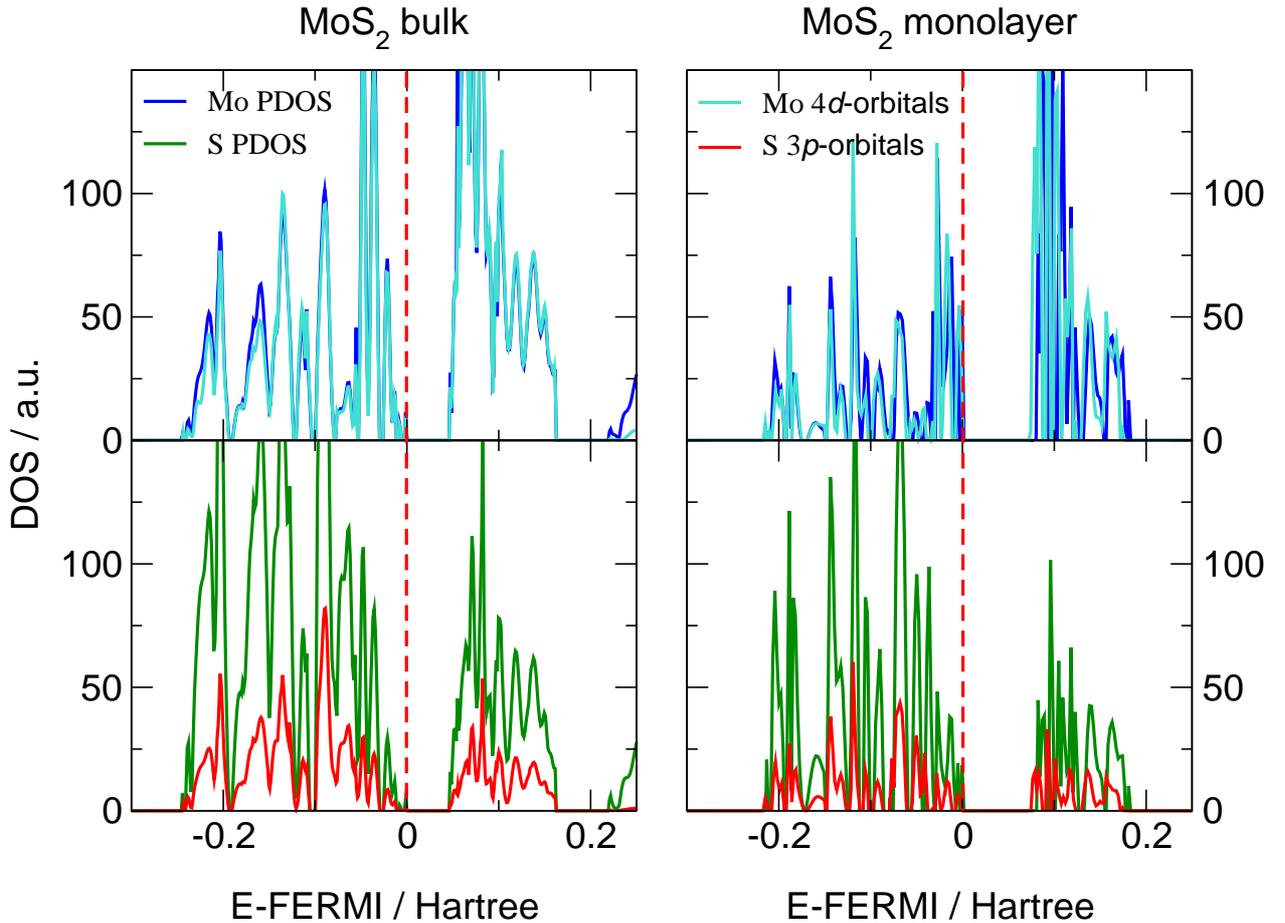}
\caption{\label{fig:5}Partial density of states of bulk MoS$_2$ and its monolayer calculated as the DFT/PBE. The projections of Mo and S atoms are given together with the contributions from 4$d$ and 3$p$ orbitals of Mo and S, respectively. The vertical dashed lines indicate the Fermi level. (Online color).}
\end{center}
\end{figure}

These unusual electronic structures of MoS$_2$ and WS$_2$, and the resulting optical properties come from the $d$-electron orbitals that dominate the valence and conduction bands (see Fig.~\ref{fig:5} for MoS$_2$).
The projected density of states (PDOS) of shows that $p$-states of sulfur atoms hybridize with the $d$-states of the transition metal atoms at the top of valence band and the bottom of conduction band.
The core states are dominated by the $s$-orbitals of the chalcogenide atom. 

\begin{figure}[ht!]
\begin{center}
\includegraphics[scale=0.60,clip]{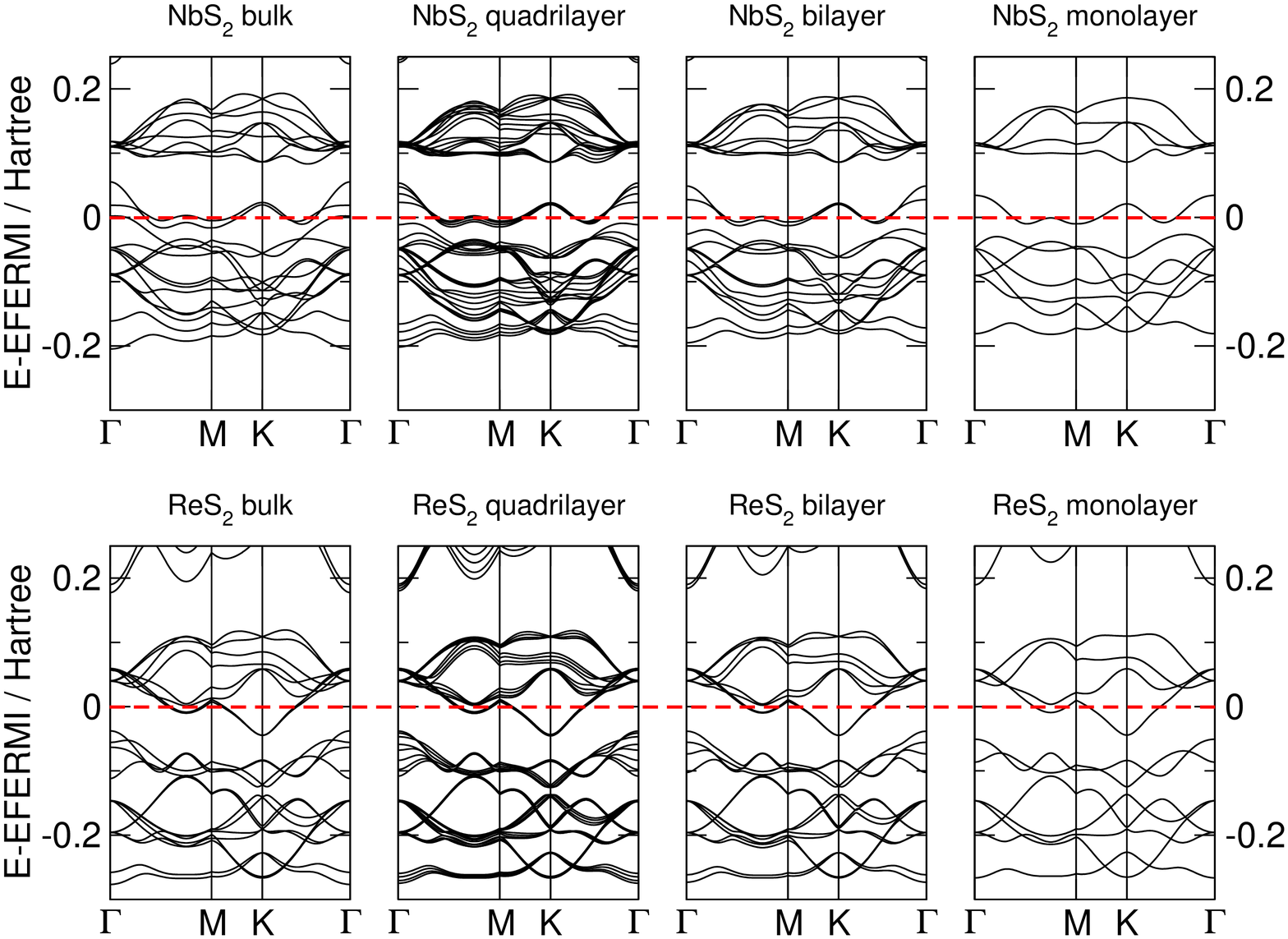}
\caption{\label{fig:6}Band structures of bulk NbS$_2$ and ReS$_2$, their monolayer, as well as, polylayers calculated at the DFT/PBE level. The horizontal dashed lines indicate the Fermi level.}
\end{center}
\end{figure}

We have also studied the role  of quantum confinement in niobium and rhenium disulfides and if their band gaps can be tuned by changing the size of the material.
The electronic band structures of NbS$_2$ and ReS$_2$ (see Fig.~\ref{fig:6}) are different from the discussed above, especially close to the Fermi level.
Although all the studied sulfides are isotopological, the NbS$_2$ and ReS$_2$ materials have metallic character independent of the number of layers.

The metallic character of NbS$_2$ comes from the fact that the $4d_{z^2}$ orbital is half-filled and results in a band that crosses the Fermi level in several points in the Brillouin zone.
The band at the Fermi level is derived primarily from these $4d$-orbitals and becomes separated from all other states in a monolayer.

The structure of ReS$_2$ studied here is of the same symmetry as all the other LTMDCs, even though the experimentally obtained material is triclinic.
ReS$_2$ has one electron per metal atom more than MoS$_2$ and has a metallic character similar to NbS$_2$.
The band that crosses the Fermi level comes from the $5d$-orbitals of Re atoms.
Unlike NbS$_2$, this band in not separated even in the monolayer and the electronic structure in the vicinity of the Fermi level is determined by short-range interactions in the sulfur $3p$ and rhenium $5d$ band complex.

For both, NbS$_2$ and ReS$_2$, the electronic structure of a bulk can be resemble by a single layer of these compounds.

\section{Conclusions}

We have studied layered transition-metal dichalcogenides in form of TmS$_2$ by means of GGA-DFT (PBE) and a hybrid functional (PBE0).
The former was found to match experimental results of the bulk phases within less than 0.1 eV, while PBE0 overestimates the band gap by 1 eV, but gives otherwise very similar band strcutures.
Our DFT, at the PBE level, support the recent findings of Splendiani et al.\cite{Splendiani2010} for MoS$_2$: the material changes its electronic properties from an indirect semiconductor in the bulk phase to a direct semiconductor in the monolayer, an interesting phenomenon with potential application for optical devices.
We show that WS$_2$ has very similar properties and can be considered as alternative material.
On the other hand, both ReS$_2$ and NbS$_2$ systems have metallic character independent of the number of layers.

The LTMDCs, if mixed together, have interesting and promising technological potential, especially for nanolelectronics and catalysis.
Therefore, we are presently working on the properties of doped structures of transition-metal dichalcogenides.

\providecommand{\refin}[1]{\\ \textbf{Referenced in:} #1}

\end{document}


\maketitle

\begin{figure}[ht!]
\begin{center}
\includegraphics[scale=0.65,clip]{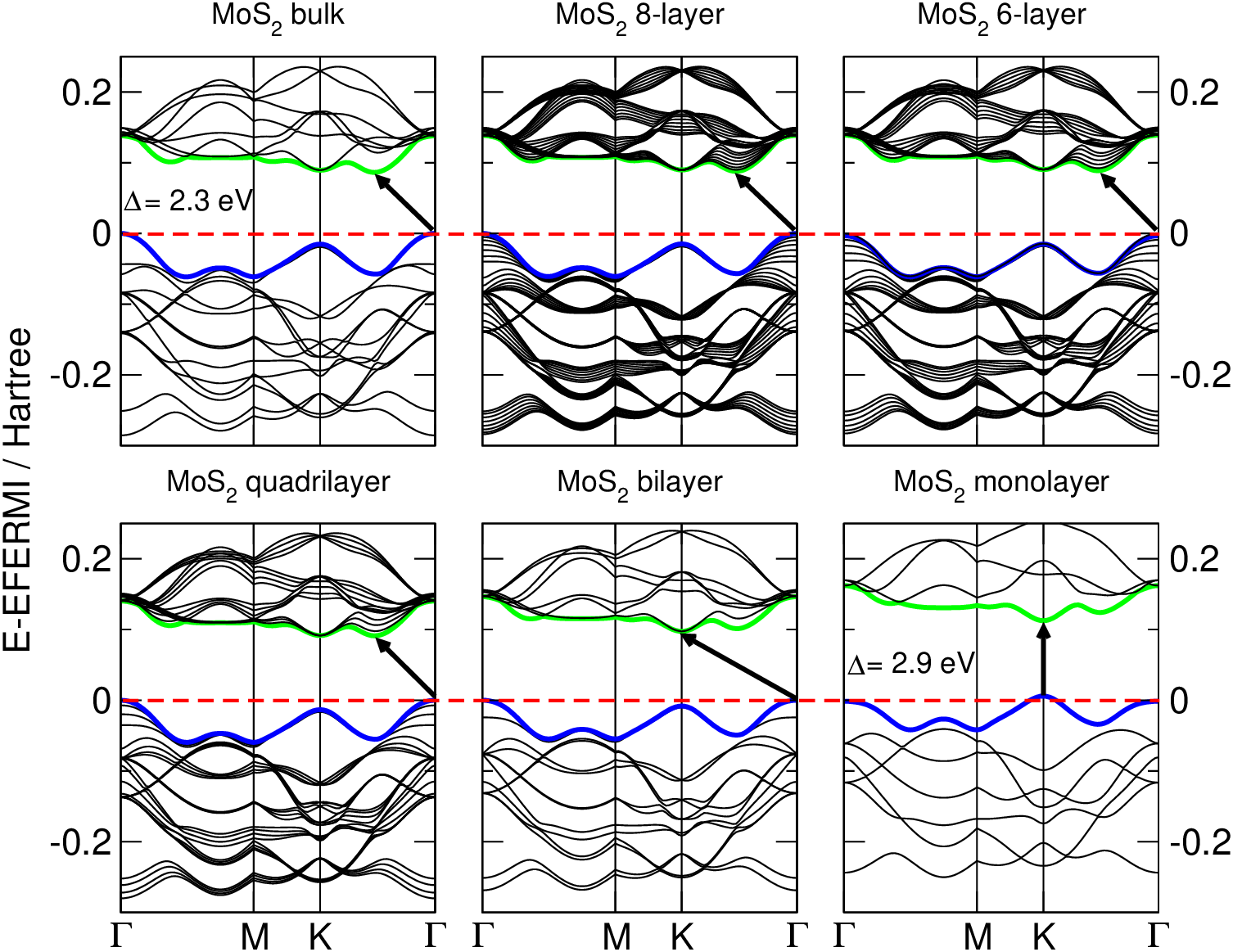}
\vspace*{-0.5cm}
\caption{\label{fig:3}\footnotesize{Band structures of bulk MoS$_2$, its monolayer, as well as, polylayers calculated as the DFT/PBE0. The horizontal dashed lines indicate the Fermi level. The arrowas indicate the smallest value of the band gap (direct or indirect) for a given system.}}
\end{center}
\end{figure}

\begin{figure}[ht!]
\begin{center}
\includegraphics[scale=0.65,clip]{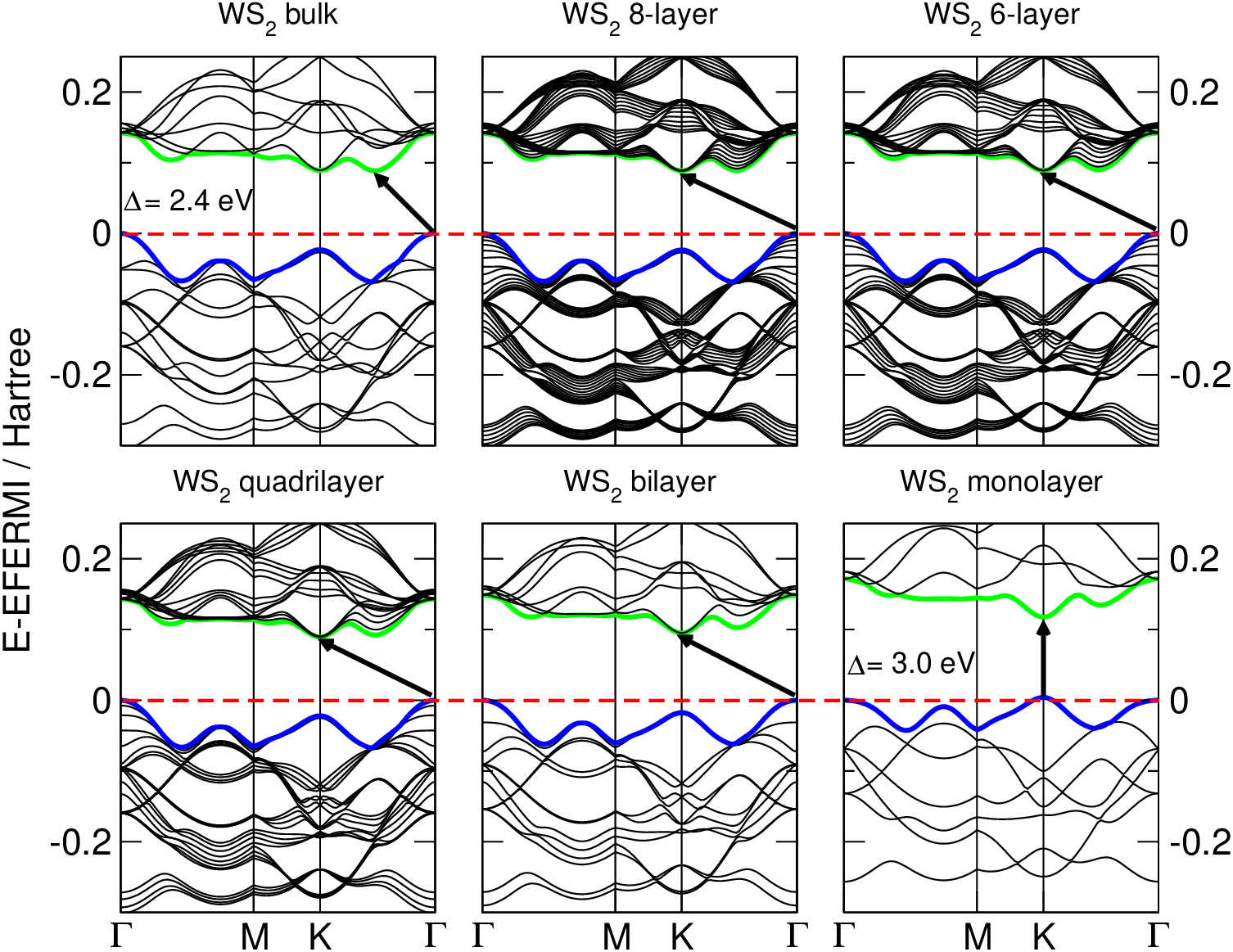}
\vspace*{-0.5cm}
\caption{\label{fig:4}\footnotesize{Band structures of bulk WS$_2$, its monolayer, as well as, polylayers calculated as the DFT/PBE0. The horizontal dashed lines indicate the Fermi level. The arrowas indicate the smallest value of the band gap (direct or indirect) for a given system.}}
\end{center}
\end{figure}

\begin{figure}[ht!]
\begin{center}
\includegraphics[scale=0.65,clip]{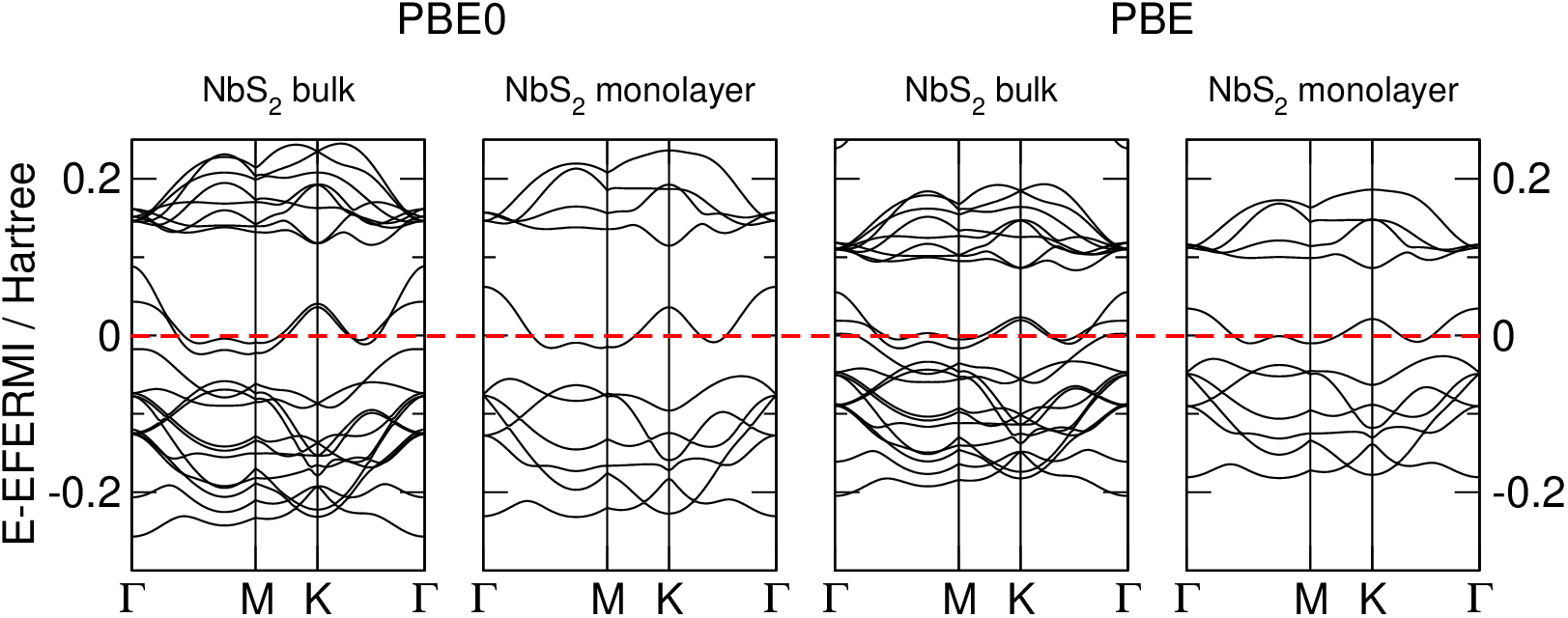}
\vspace*{-0.5cm}
\caption{\label{fig:7}\footnotesize{Band structures of bulk NbS$_2$ and its monolayer calculated as the DFT/PBE0 and the DFT/PBE. The horizontal dashed lines indicate the Fermi level. The arrowas indicate the smallest value of the band gap (direct or indirect) for a given system.}}
\end{center}
\end{figure}

\begin{figure}[ht!]
\begin{center}
\includegraphics[scale=0.65,clip]{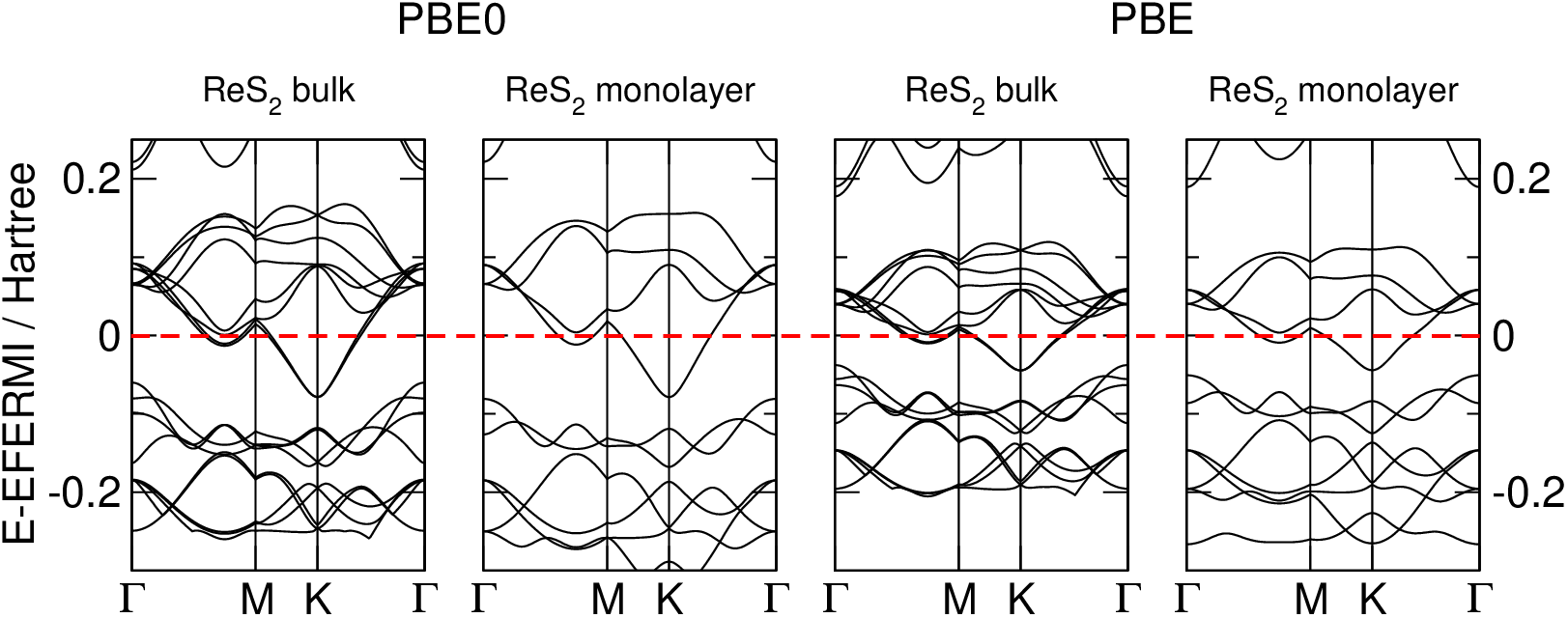}
\vspace*{-0.5cm}
\caption{\label{fig:8}\footnotesize{Band structures of bulk ReS$_2$ and its monolayer calculated as the DFT/PBE0 and the DFT/PBE. The horizontal dashed lines indicate the Fermi level. The arrowas indicate the smallest value of the band gap (direct or indirect) for a given system.}}
\end{center}
\end{figure}

\begin{figure}[ht!]
\begin{center}
\includegraphics[scale=0.65,clip]{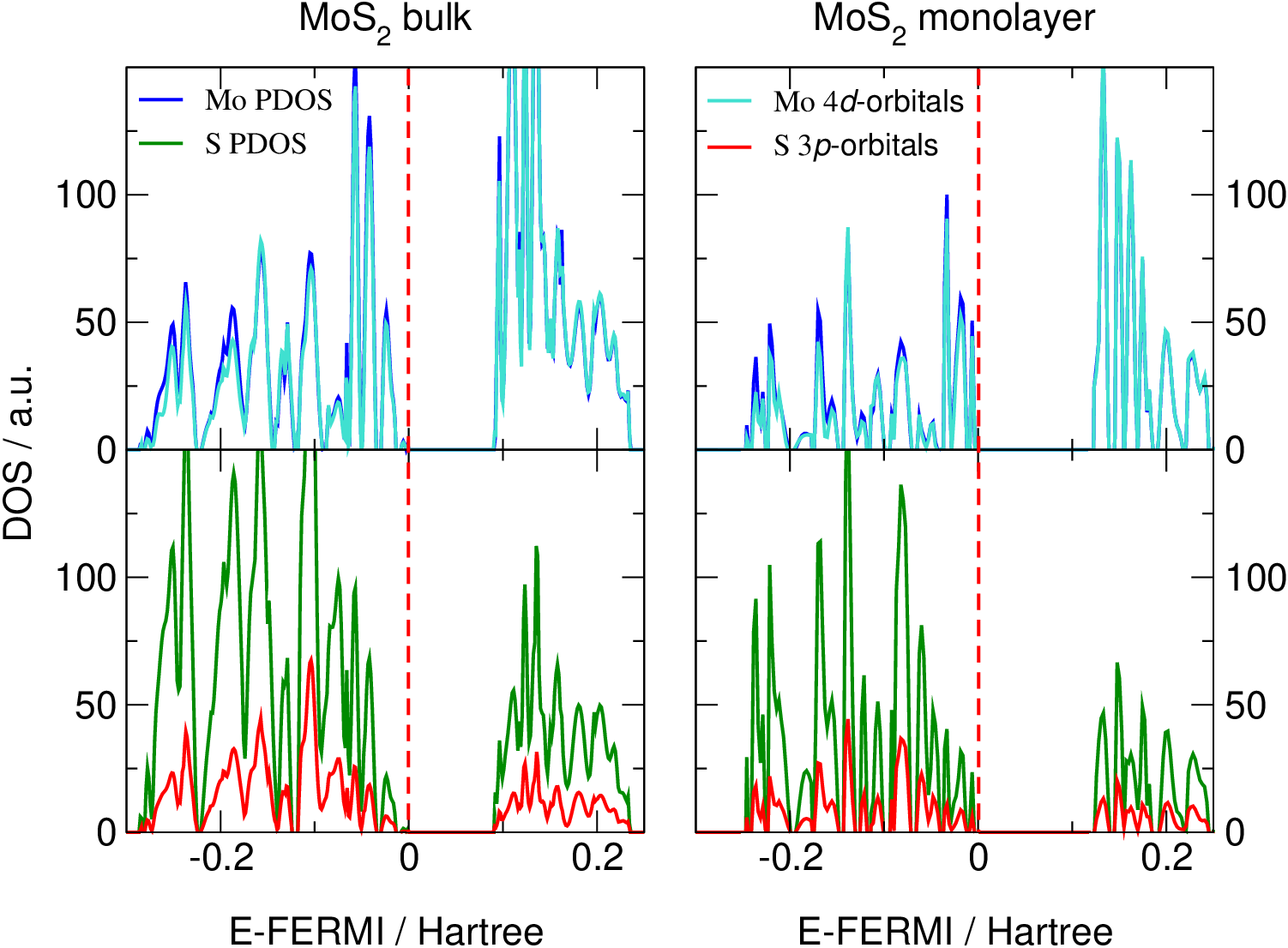}
\caption{\label{fig:5}Partial density of states of bulk MoS$_2$ and its monolayer calculated as the DFT/PBE0. The projections of Mo and S atoms are given together with the contributions from 4$d$ and 3$p$ orbitals of Mo and S, respectively. The vertical dashed lines indicate the Fermi level. (Online color).}
\end{center}
\end{figure}